\documentclass{article}

\usepackage{arxiv}

\usepackage[utf8]{inputenc} 
\usepackage[T1]{fontenc}    
\usepackage{hyperref}       
\usepackage{url}            
\usepackage{booktabs}       
\usepackage{amsfonts}       
\usepackage{nicefrac}       
\usepackage{microtype}      
\usepackage{lipsum}		
\usepackage{graphicx}
\usepackage{natbib}
\usepackage{doi}
\usepackage{amsmath}

\title{"The Simplest Idea One Can Have" for Seamless Forecasts with Postprocessing}

\author{ \href{https://orcid.org/0000-0002-2238-4282}{\includegraphics[scale=0.06]{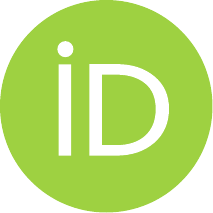}\hspace{1mm}Markus~Dabernig}\\
	Analysis and Model Development\\
	GeoSphere Austria\\
	Hohe Warte 38, 1190 Vienna, Austria \\
	\texttt{markus.dabernig@geosphere.at} \\
	\And
	\href{https://orcid.org/0000-0002-4174-4335}{\includegraphics[scale=0.06]{orcid.pdf}\hspace{1mm}Aitor~Atencia} \\
	Analysis and Model Development\\
	GeoSphere Austria\\
	Hohe Warte 38, 1190 Vienna, Austria \\
	\texttt{aitor.atencia@geosphere.at} \\
}

\renewcommand{\headeright}{ }
\renewcommand{\undertitle}{ }
\renewcommand{\shorttitle}{"The Simplest Idea One Can Have"}

\hypersetup{
pdftitle={The Simplest Idea One Can Have for Seamless Forecasts with Postprocessing},
pdfauthor={Markus~Dabernig, Aitor~Atencia},
pdfkeywords={Seamless, Postprocessing, Persistence, Multimodel},
}

\begin{document}
\maketitle

\begin{abstract}

Seamless forecasts are based on a combination of different sources to produce the best possible forecasts. Statistical multimodel postprocessing helps to combine  various sources to achieve these seamless forecasts. However, when one of the combined sources of the forecast is not available due to reaching the end of its forecasting horizon, forecasts can be temporally inconsistent and sudden drops in skill can be observed. To obtain a seamless forecast, the output of multimodel postprocessing is often blended across these transitions, although this unnecessarily worsens the forecasts immediately before the transition. Additionally, large differences between the latest observation and the first forecasts can be present. 

This paper presents an idea to preserve a smooth temporal prediction until the end of the forecast range and increase its predictability. This optimal seamless forecast is simply accomplished by not excluding any model from the multimodel by using the latest possible lead time as model persistence into the future. Furthermore, the gap between the latest available observation and the first model step is seamlessly closed with the persistence of the observation by using the latest observation as additional predictor. With this idea, no visible jump in forecasts is observed and the verification presents a seamless quality in terms of scores. The benefit of accounting for observation and forecast persistence in multimodel postprocessing is illustrated using a simple temperature example with linear regression but can also be extended to other predictors and postprocessing methods.
 
\end{abstract}


\section{Introduction}
Numerical weather predictions (NWP) are expensive to calculate, especially high-resolution models on a limited area, and are therefore usually restricted to shorter lead times \citep{Wastl2021}. In contrast, global models, for example, the European Centre for Medium-Range Weather Forecasts (ECMWF), predict further into the future but on a coarser grid. Postprocessing methods can be applied to achieve the optimal combination of these different models. However, usually, a jump in performance is present at the transition from the shorter to the more extended model as shown in \citet{Keller2021}, \citet{Rust2023} or \citet{Vannitsem2021}. Linear blending can be applied to overcome this jump by arbitrarily reducing the skill of the high-resolved model in favour of the long-term range models. A second weakness of NWP is that the initial state is far from the latest available observation for different reasons, such as model stability during the data assimilation process but also the runtime and the resulting delay of availability (up to several hour). Therefore, any seamless forecast using only a combination of NWP models starts with a large error at the first lead time. 
This paper proposes an idea to overcome both problems and to maintain the best possible combination at any time. The idea is to use observation persistence to close the performance gaps between the latest observation and the first lead time. The same principle is applied to shorter forecast horizon models by using the latest available lead-time forecast as a persistence afterwards. The persistence to close the gap to the latest observations is using precisely this latest observation as an additional predictor for all lead times of the multimodel postprocessing. The so-called persistence to close the gap between models is the last available forecast from the shorter model as an additional predictor for all following lead times. Therefore, the same number of predictors (models and persistences) are used in all the lead times, and it is possible to close these gaps and maintain the best possible predictions without artificially modifying the weighting ass in a linear blending approach. This paper is not intended to give an exhaustive investigation but to communicate this simple idea to produce seamless forecasts. In the following, we will present a simple postprocessing method and its verification at a few stations as a proof of concept.

\section{Data}

The experiments are based on temperature forecasts of the local model AROME, ECMWF~deterministic and ECMWF~ensemble mean. To better show this idea's concept, AROME is cut at +36~h lead time and ECMWF deterministic at +84~h lead time, although more extended forecast periods are available. Until lead time +84~h, the models are in hourly resolution and 3 hourly after. The postprocessing is initialised at 12~UTC, while all models are initialised at 00~UTC. All lead times refer to the postprocessing initialisation and not the NWP initialisations. The forecasting horizon for the seamless forecast is 132 hours.
Observations from the automatic weather stations network over Austria from GeoSphere Austria on three different locations are selected. The locations are chosen with distinct characteristics. One station is in a plain in Vienna where observations and NWP are roughly the same height. Another station is in a valley (Innsbruck) where the NWP height is higher than the height of the weather station. The third and last station is on a mountain (Sonnblick) where the NWP height is lower than the actual height.
The data used comprises the period from July 2021 until June 2024 and is with 3-fold cross validation split into two years training and one year testing respectively.

\section{Method}
Ensemble model output statistics (EMOS), as introduced in~\citet{Gneiting2005}, is the selected postprocessing method. EMOS with a smooth function to adjust to a seasonal bias but as a homoscedastic model with a constant scale parameter as presented in \citet{Demaeyer2023} is used in this study. The homoscedastic model was chosen for simplicity, but a heteroscedastic model could also be used, obtaining similar performance. The model is formulated with the following equations where the temperature is represented by a Gaussian distribution where the location is modelled with a linear regression, and the scale is a constant: 

\begin{equation}
y \sim  \mathcal{N} ( \mu, \text{const.}) 
\label{eq1}
\end{equation}

\begin{equation}
\mu = \beta_0 + \beta_1 \text{pers} + \beta_2 \text{aro} + \beta_3 \text{det} + \beta_4 \text{ens}_\mu + f_1(\mathrm{doy})
\label{eq2}
\end{equation}

with $\beta_0$ to $\beta_4$ as regression coefficients,  \text{pers} as the persistence of the observation,  \text{aro} as AROME,  \text{det} as ECMWF~deterministic and $\text{ens}_\mu$ as ECMWF~ensemble mean forecast. $f_1(\mathrm{doy})$ is a combination of base functions of the day of the year ($\sin(2\pi\,\text{doy}/365)$,\, $\cos(2\pi\,\text{doy}/365)$,\, $\sin(4\pi\,\text{doy}/365)$, and, $\cos(4\pi\,\text{doy}/365)$) as presented in~\citet{Dabernig2017} to capture the seasonal bias. The implemented EMOS version is based on the R-package \texttt{crch}~\citep{Messner2016} with maximum likelihood estimation.  

The basic idea to close gaps after, for example, AROME is not available anymore is to keep the last forecast as a predictor for all following lead times. For example, at lead time +40~h, ECMWF deterministic and ensemble mean are used from +40~h, but AROME is taken from +36~h. Similarly, to close the gap to observation at lead time +1~h, the observation from +0~h is used as a predictor (and also at lead time +40~h). As proof of concept, we also use the observation from +0~h as a predictor at the lead time +132~h, which is probably optional. At the last transition at +84~h lead time, the deterministic forecast from +84~h is kept as a predictor until +132~h lead time. As a result the multimodel has the same amount of predictors at lead time +1~h as at lead time +132~h.

The reference would be a multimodel of all available NWP as input, but it reduces to only ECMWF deterministic and ensemble after +36~h and to only ECMWF ensemble after +84~h. The observation persistence is not used in the reference model either. Additionally for comparison, EMOS was also applied to every individual model. Therefore, Equation~\ref{eq2} reduces to an intercept ($\beta_0$), the seasonal smooth term $f_1(\mathrm{doy})$ and the corresponding model term ($\beta_2 \text{aro}$, $\beta_3 \text{det}$ or $\beta_4 \text{ens}_\mu$).

\section{Results}
The mean absolute error (MAE) was calculated at the three stations to show the effect of the presented idea. Additionally, the MAE skill score in percent ($(1 - \text{MAE} / \text{MAE}_\text{ref}) * 100 $) is shown to demonstrate the improvement compared to the reference model.
\begin{figure}[bt]
\centering
\includegraphics[width=\textwidth]{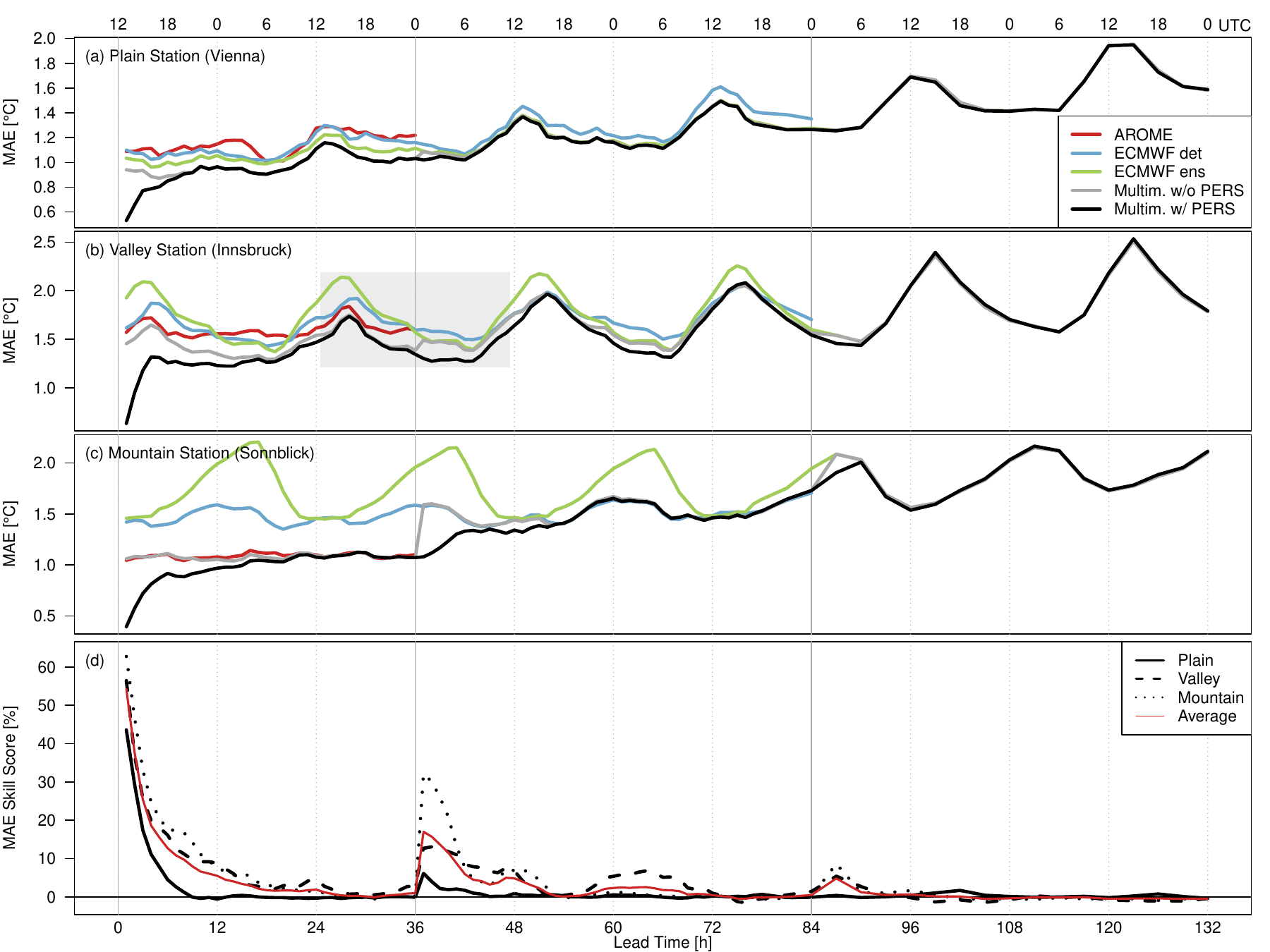}
\caption{Mean absolute error at three stations ((a) plain station, (b) valley station and (c) mountain stations) for the forecasts with a single NWP as input (colours), multimodel without persistence (grey) as a reference model and multimodel with persistence (black). Forecasts are between lead times +0 and +132~h. Vertical lines at +36 and +84~h indicate a transition where a model disappears, and persistence was used as a predictor. Grey shading in (b) shows the detailed area for Figure~\ref{fig:verif_detail}. (d) shows the MAE skill score of the multimodel with persistence using the multimodel without persistence as a reference at the three different stations (black lines) and an average of all stations (red line).}
\label{fig:verif}
\end{figure}

Figure~\ref{fig:verif} shows a large improvement in the first few lead times at all three stations. At lead time +1~h an improvement of skill between 45 and 65~$\%$ can be achieved by using the latest available observation as additional predictor. This improvement is even present up to almost +24h at the mountain and valley station (a small peak in skill score is visible which shows and improvement of performance based on the observation persistence), while the plain stations show only an influence of the persistence up to +12~h. 
After +36~h when only ECMWF deterministic and ensemble is available, the model persistence of AROME can improve the multi model with only a few percent points at the plain station, over 10 percent points at the valley station and up to 30~percent points at the mountain station. However, these improvements last shorter as compared to the observation persistence but seem to return at the valley station between +60~h and +70~h. Therefore, to not arbitrary reduce the skill of the persistence we suggest to keep the persistence until the last lead time and use variable selection methods such as gradient boosting (\citet{Messner2017}) to guarantee the best possible forecast.
At the last transition at +84~h the improvements are much smaller since ECMWF deterministic has an already reduced skill compared to ECMWF ensemble. However, at the valley and mountain station the model persistence of ECMWF deterministic still improves the combination between 5 and 10~percent points.

\begin{figure}[bt]
\centering
\includegraphics[width=0.45\textwidth]{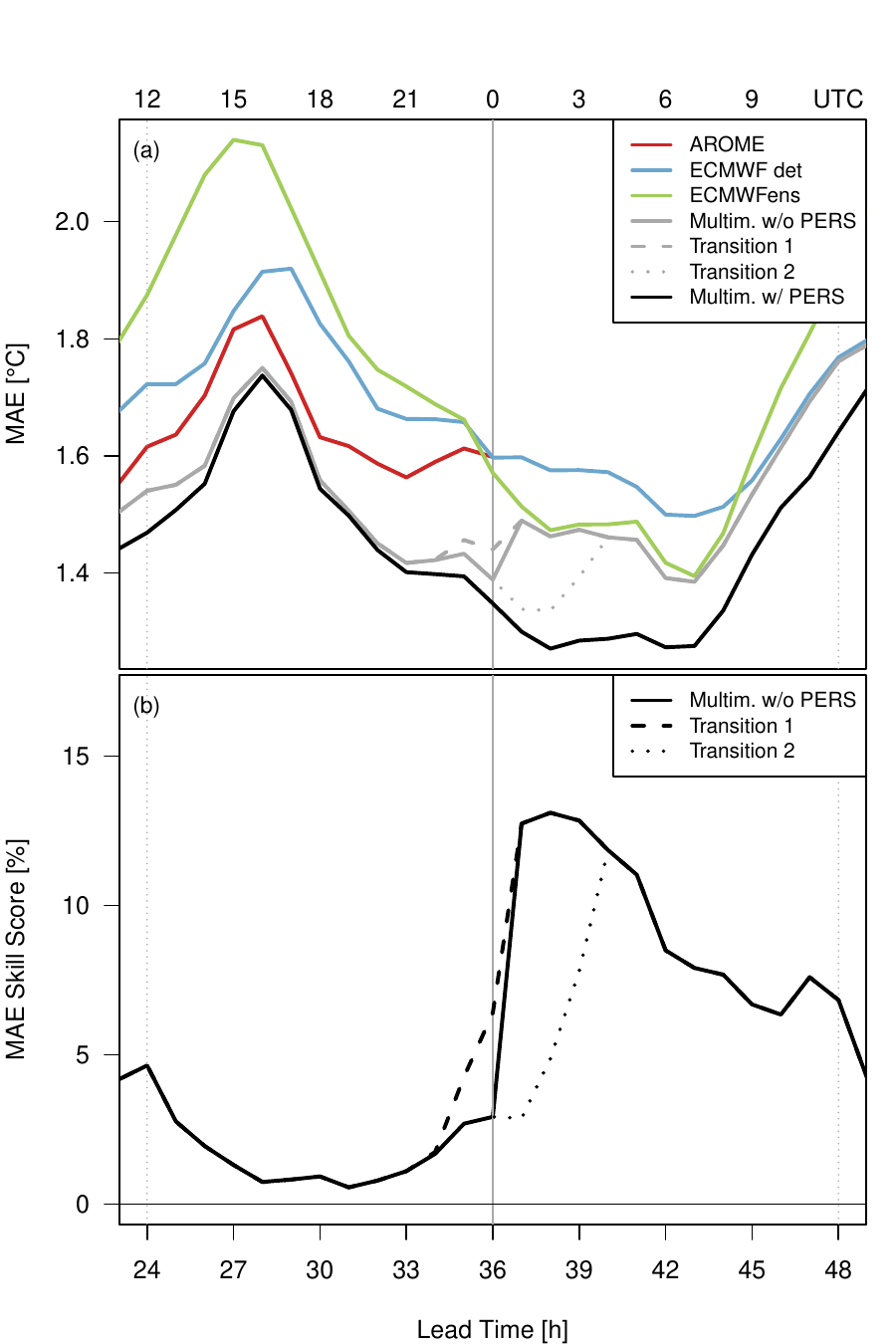}
\caption{(a) Mean absolute error at valley stations as in Figure~\ref{fig:verif}b but only for the grey shaded area with 2 additional transitions (grey dashed and grey dotted) as suggested in \citet{Keller2021}. (b) MAE Skill Score as in Figure~\ref{fig:verif}d but additional showing the performance of the multimodel with persistence compared to transition 1 and transition 2 as reference.}
\label{fig:verif_detail}
\end{figure}

Figure~\ref{fig:verif_detail} provides a closer look at the verification for the valley station at lead time +36~h where the local model ends and only two global models are available. Additionally to the methods plotted in Figure~\ref{fig:verif}, two transition methods are shown as suggested in \cite{Keller2021}. These two transition methods for seamless forecasting are: the "weighted average" method (transition 1), which combines forecasts from different models using time-dependent weights, and the  "extrapolation" method (transition 2), which selects the latest post-processed result  as a weighted member for latter time steps. These methods aim to create smooth transitions between different forecast models to improve temperature predictions in areas with complex topography. While both transitions are used to achieve a linear blending from with to without local model, an evident loss of performance compared to the model with persistence is present. Transition 1 already reduces the skill before lead time +36~h, while AROME is still available. Transition 2 is able to increase the skill a little further into the future even after lead time +36~h but only for three lead times (as constructed). Therefore, using the latest available model persistence outperforms not only a multimodel without persistence but also the techniques to achieve seamless transitions, such as linear blending.  

In summary, the observation and model persistence do not always improve the combination. Still, it substantially improves performance in certain stations without any equivalent worsening in the "worse" lead times. Yet, the most crucial feature is that the introduction of these persistences produces a clearly seamless forecast (as can be observed in Figure~\ref{fig:verif} a,b and c) without visible jumps around the selected times (horizontal grey solid lines) in the MAE. It also eliminates the need to make an arbitrary/linear transition, reducing the seamless forecast's skill even further (as in Figure~\ref{fig:verif_detail}).  

\section{Conclusion}
The presented method in this paper proves "the simplest idea one can have   "\footnote{The reaction we got when first presenting this idea.} to produce optimal seamless multimodel forecasts with the best possible skill without jumpiness (not seamless) and without predefined blending weights when missing predictors (not optimal) by just using the persistence of the latest observation and the latest available lead time of a model. This idea leads to substantial improvements, especially in the first lead times to close the gap between NWP and observations, as well as at the end of a model range. Hence, the transition from the local model to the global model is improved. Not all stations show the same magnitude of improvement, however using the several persistence in the different types of stations used in this study does not have any considerable disadvantage. While the presented postprocessing only has one predictor per model as input, experience showed that similar results can be achieved by using the persistence of all available predictors of every model. However, a variable selection method would be needed to avoid a general overfitting \citep{Messner2017}. 

The idea of using observation and the latest lead-time model as persistence is also only shown for temperature. Still, experiments with other variables such as wind, gusts, relative humidity, and also precipitation achieve similar results. On the same note, although only a homoscedastic example was shown, other investigations have shown that the scale parameter of the distribution behaves similarly to the location of the distribution and is adjusted to correspond to the improvement with persistence. Consequently, the conclusions can be extended to other variables and when heteroscedasticity is introduced.

While the presented method is only shown on postprocessing at stations (EMOS), it is also used on spatial forecasts (SAMOS in \citet{Dabernig2017, Dabernig2020}) in an operational setting and it achieves similar improvements, especially at the beginning but also largely at the transition from local to global model. 
Even if the idea was presented with a simple postprocessing method, it could be beneficial for any postprocessing method which is applied lead time-wise, as, for example, several methods presented in \citet{Vannitsem2021} and \citet{Demaeyer2023}.

An additional advantage of the method is that even if the NWP is not initialised every hour, improved short-term forecasts can be recalculated every hour with the latest available observations. This brings a huge advantage, especially for wind power or air quality forecasts and nowcasting in general, since it is computationally cheap and substantially improves the first few lead times. 

\section*{Acknowledgements}
The authors thank GeoSphere Austria (former Zentralanstalt für Meteorologie und Geodynamik) for providing the automatic weather station (AWS) information from different sources such as TAWES, etc. The authors would also thank the NWP Competence Unit for providing valuable access to different NWP models used in this study. This research is funded by GeoSphere Austria through the Entwicklungsprojekt "Nowcasting Strategy - Year 1" (23-BM-EWP-013). Also thanks to Thomas Muschinski for an internal revision and his suggestions to improve to this paper.

\bibliographystyle{plainnat_V2}
\bibliography{references}  

\end{document}